\documentclass[12pt, a4paper]{article}
\usepackage{cite}
\usepackage{amsmath,amssymb}
\usepackage{comment}
\usepackage{bm}
\usepackage{braket}
\usepackage{url}
\usepackage{accents}
\usepackage{slashed}
\usepackage{utfsym}
\usepackage{tikz-feynman}
\usepackage{multirow}

\usepackage{ifpdf}
\ifpdf
  \usepackage{graphicx, hyperref, xcolor}     
\else     
  \usepackage[dvipdfmx]{graphicx, hyperref, xcolor}     
\fi

\definecolor{rossoferrari}{HTML}{D9073D}
\definecolor{mediumblue}{HTML}{0000CD}
\hypersetup{
setpagesize=false,
bookmarksnumbered=true,%
bookmarksopen=true,%
colorlinks=true,%
linkcolor=rossoferrari,
urlcolor=mediumblue,
citecolor=mediumblue,
}

\newcommand{\tmin}{t^{\mathrm{(min)}}}
\newcommand{\tmax}{t^{\mathrm{(max)}}}
\newcommand{\nstep}{N_{\text{step}}}
\newcommand{\nshot}{N_{\text{shot}}}
\mathchardef\mhyphen="2D

\begin{document}

\begin{titlepage}

\begin{center}

\hfill KEK-TH-2417\\

\vskip .75in

{\Large \bf Quantum Simulations of Dark Sector Showers}

\vskip .75in

{\large So Chigusa$^{\usym{1F019}, \usym{1F01A}, \usym{1F01B}}$ and Masahito Yamazaki$^{\usym{1F01C}}$}

\vskip 0.25in

\usym{1F019} {\em Berkeley Center for Theoretical Physics, Department of Physics,\\
University of California, Berkeley, CA 94720, USA}\\[.3em]

\usym{1F01A} {\em Theoretical Physics Group, Lawrence Berkeley National Laboratory,\\
Berkeley, CA 94720, USA}\\[.3em]

\usym{1F01B} {\em KEK Theory Center, IPNS, KEK, Tsukuba, Ibaraki 305-0801, Japan}\\[.3em]

\usym{1F01C} {\em Kavli Institute for the Physics and Mathematics of the Universe (WPI),\\
University of Tokyo, Kashiwa, Chiba 277-8583, Japan}

\end{center}
\vskip .5in

\begin{abstract}
We consider dark sector scenarios where dark matter is accompanied by a dark photon and multiple-flavor dark fermions charged under the dark gauge group. We study quantum interference effects in dark sector jets, where multiple dark photons are emitted from high-energy dark fermions. We perform fully quantum simulations of dark sector showers and compare the results against those of the classical Monte-Carlo simulations. We find important differences in probability distributions of dark photon countings between quantum and classical computations. When the number of dark-fermion flavors is large, we find significant enhancements in large numbers of dark photon emissions. Such enhancements can provide distinguishing signals for our scenarios at particle colliders.

\end{abstract}

\end{titlepage}

\renewcommand{\thepage}{\arabic{page}}
\setcounter{page}{1}
\renewcommand{\thefootnote}{\#\arabic{footnote}}
\setcounter{footnote}{0}
\renewcommand{\theequation}{\thesection.\arabic{equation}}


\section{Introduction}
\setcounter{equation}{0}

The existence of dark matter (DM) is one of the most important hints of physics beyond the Standard Model (SM).
While numerous DM models have been proposed by many authors,
all the observational evidence of DM is based on its gravitational interaction, and we have only limited knowledge of its mass or other interactions. 

In recent years, several cosmological experiments have favored self-interacting dark matter (SIDM) models.
For example, the observed large-scale structure of the universe might be incompatible with the predictions of traditional cold DMs \cite{Spergel:1999mh}, and this discrepancy can be resolved if we introduce SIDM with a self-interaction cross section normalized by its mass $\sigma/m_\chi \sim 0.1 \mhyphen 10\,\mathrm{cm^2/g}$ \cite{vogelsberger2012subhaloes, Rocha:2012jg, zavala2013constraining}.
These models can also explain the positron excess in the cosmic ray observed by, e.g., PAMELA, FERMI-LAT, and AMS-02 \cite{PAMELA:2008gwm, adriani2010statistical, ackermann2012measurement, AMS:2019rhg}.
If we consider a scenario where DM annihilates into mediators of the DM self-interaction, the observed positron spectrum can be explained by the mediator decay, while the anti-proton spectrum can be kept intact if the decay into a proton-antiproton pair is forbidden by the kinematics \cite{Pospelov:2007mp, Cholis:2008vb, Pospelov:2008jd, Nelson:2008hj, Cholis:2008qq, Feldman:2008xs, Bergstrom:2008ag, Arkani-Hamed:2008hhe}.
Also, the SIDM models can explain the Galactic Center GeV excess observed by FERMI-LAT \cite{Goodenough:2009gk, vitale2009indirect, Hooper:2010mq, Hooper:2011ti, Abazajian:2012pn, Daylan:2014rsa, Fermi-LAT:2015sau}.
In general, models that explain the excess by direct annihilations of DM into the SM particles suffer from severe constraints from DM direct detection.
However, the DM annihilation into dark mediators allows us to construct successful models \cite{Hooper:2012cw, Martin:2014sxa, Boehm:2014bia, Abdullah:2014lla, Berlin:2014pya, Cline:2014dwa, Liu:2014cma, Cline:2015qha, Elor:2015tva}.

The unknown part of the SIDM models, including the DM and the dark mediators, is often called the dark sector.
Among the possibilities for dark sectors, arguably the simplest is to introduce a new Abelian gauge symmetry $U(1)_D$ and a fermionic DM $\chi$ charged under it.
The dark mediator in this model is the so-called dark photon $A'$.
The dark photon can have a tiny kinematic mixing with the gauge boson of the SM hypercharge $U(1)_Y$, through which the dark photon decays into SM particles and the dark sector constituents interact with their SM counterparts.

Signals originating from the dark sector have been explored at collider experiments.
When the self-interaction of dark matter is strong (as suggested for example by the large-scale structure of the universe), the dark sector shower can generate one of the most characteristic signals at the colliders.
After dark sector particles are first created at the collider, a large number of dark particles are produced through the emission of $A'$s from $\chi$ and the pair production of $\chi$s by $A'$, similar to the parton shower in QCD.
The dark sector shower may be converted back into Standard Model particles through, e.g., the decay of dark photons.
These events are efficiently searched for by utilizing displaced signals \cite{Schwaller:2015gea, Pierce:2017taw, Renner:2018fhh, Bernreuther:2020xus, Yuan:2020eeu}, lepton jets \cite{Cheung:2009su, Meade:2009rb, Falkowski:2010cm, Falkowski:2010gv},
jets/event shapes \cite{Harnik:2008ax, Cohen:2015toa, Kim:2016fdv, Knapen:2016hky, Cohen:2017pzm, Beauchesne:2017yhh, Park:2017rfb, Heimel:2018mkt, Cohen:2020afv, Bernreuther:2020vhm}, and so on.

Classical parton shower calculation works well to simulate the dark sector showers in many cases, but it misses some of the crucial quantum interference effects (see, e.g., \cite{DOKSHITZER1980269}).
This is particularly the case if $\chi$ has a non-trivial flavor structure, 
where interference between diagrams that contain different flavors of intermediate state fermions is not properly taken into account.
Recently, several algorithms are developed \cite{Bauer:2019qxa, Bepari:2020xqi, Li:2021kcs, Williams:2021lvr, Macaluso:2021ngq} that correctly incorporate the quantum effects and can be implemented on a quantum computer.
In this paper, we follow \cite{Bauer:2019qxa} and quantify the impact of quantum interference effects on the calculation of dark parton showers containing two flavors of $\chi$s as a simple example. 
There is an advantage in studying quantum parton showers in the context of dark sector models---simple models, which are meant to be toy models for QCD in \cite{Bauer:2019qxa}, can describe the actual dark sector model searched in colliders. We compare the results of classical and quantum calculations in the expectation values and the variances of the number of dark photon emissions.

The organization of the paper is as follows. 
In Section \ref{sec:dark-shower}, we review classical calculations of dark sector showers.
In Section \ref{sec:interference}, we evaluate quantum interference effects on the dark sector showers and obtain analytic expressions of the effects for a specific choice of model parameters.
In Section \ref{sec:simulation}, we compare classical and quantum calculations under various model parameters and 
discuss when the quantum effects are important.
In Section \ref{sec:discussion}, we conclude this paper with further discussions on the implications of our results.

\section{Review of classical dark sector showers}
\setcounter{equation}{0}
\label{sec:dark-shower}

In this section, we review the classical calculation of dark sector shower, following the standard treatment in the literature (see, e.g., Ref.~\cite{Buschmann:2015awa}).
We here concentrate on the model with only one flavor of the dark fermion.
The multi-flavor case is discussed in the next section.

We consider the dark sector Lagrangian with the dark $U(1)_D$ symmetry
\begin{align}
  \mathcal{L}_{\mathrm{dark}} =
  \bar{\chi} (i\slashed{\partial} - m_\chi + ig \slashed{A}') \chi
  - \frac{1}{4} F'_{\mu\nu} F'^{\mu\nu} + \frac{1}{2} m_{A'}^2 A'_\mu A'^\mu\;,
\end{align}
where $A'$ is a dark photon with mass $m_{A'}$, and $\chi$ is a dark fermion whose coupling to the dark photon is determined by $g$. The field strength of the dark photon is denoted by $F'_{\mu\nu}$.
In this model, a propagating dark fermion $\chi$ experiences repeated emissions of dark photons through 
the channel $\chi \to \chi + A'$, in particular when the initial $\chi$ is hard and/or the coupling $g$ is large.
We assume $m_{A'} < m_\chi$ for simplicity and neglect the pair production process $A' \to \bar{\chi} \chi$.

Let us consider the splitting process $\chi(p_{\chi,\mathrm{in}}) \to \chi(p_{\chi,\mathrm{out}}) + A'_\mu(k)$, where the quantities in the parentheses denote the four-momenta of the incoming/outgoing dark fermions and the dark photon.
We work under the so-called collinear approximation, where all internal partons are treated to be on-shell. This approximation is valid only if the energy of the incoming parton $E$ is much larger than the parton masses, i.e., $E\gg m_\chi,\,m_{A'}$, so that every splitting is collinear.
By choosing an appropriate coordinate system, the relevant four-momenta can be expressed as
\begin{align}
\begin{split}
  p_{\chi,\mathrm{in}} &= \left(E, 0, 0, p \right)\;,\\
  p_{\chi,\mathrm{out}} &= \left( xE, -p_t, 0, \sqrt{x^2 E^2 - p_t^2 - m_\chi^2} \right)\;,\\
  k &= \left( (1-x)E, p_t, 0, \sqrt{(1-x)^2 E^2 - p_t^2 - m_{A'}^2} \right)\;,
\end{split}
\end{align}
where $p_t$ denotes the transverse momentum of the emitted dark photon and
$x$ denotes the energy fraction of the outgoing parton with respect to the incoming parton.
For the four-momenta to be real, $p_t$ and $x$ should satisfy
\begin{align}
  \frac{\sqrt{p_t^2 + m_\chi^2}}{E} \leq x \leq 1-\frac{\sqrt{p_t^2 + m_{A'}^2}}{E} \;.
  \label{eq:xrange}
\end{align}

For our purposes, it is useful to replace the transverse momentum $p_t$ by the virtuality $t$ defined as
\begin{align}
  t \equiv (p_{\chi,\mathrm{out}} + k)^2 - m_\chi^2\;.
\end{align}
The virtuality takes the minimum value $t = \tmin$ when $p_t = 0$, leading to
\begin{align}
  \tmin = m_{A'}^2 + 2m_{A'} m_\chi\;.
\end{align}
The maximum value $t = \tmax$ is obtained for the maximum value of $p_t = p_\tmax$.
A simple calculation shows $p_\tmax \simeq (E^2 + m_\chi^2 - m_{A'}^2)/(2E^2)$,
which gives
\begin{align}
\tmax = E^2 - m_\chi^2\;.
\end{align}

For a fixed value of $t$ in the region $\tmin \le t \le \tmax$, the condition \eqref{eq:xrange} can be rewritten as
\begin{align}
  x_{\mathrm{min}}(t, E^2)
  \leq x \leq
  x_{\mathrm{max}}(t, E^2)\;,
  \label{eq:xrange-int}
\end{align}
with
\begin{align}
\begin{split}
  x_{\mathrm{min}}(t, E^2) &= \max \left[ x_{-}(t,E^2), \frac{t + 2m_\chi^2 - m_{A'}^2}{2E^2} \right]\;,\\
  x_{\mathrm{max}}(t, E^2) &= \min \left[ x_{+}(t,E^2), 1- \frac{t + m_{A'}^2}{2E^2}, 1-\frac{m_{A'}}{E} \right]\;,\\
  x_{\pm}(t,E^2) &=
  \frac{E (t + 2m_\chi^2 - m_{A'}^2) \pm \sqrt{(E^2 - m_\chi^2 - t)(t^2 - 2 m_{A'}^2 (t + 2m_\chi^2) + m_{A'}^4)}}{2E (t + m_\chi^2)}\;.
 \end{split}
\end{align}

The dark photon emission can be considered a classical Markov process,
with the emission probability density $R(t)$ given by
\begin{align}\label{R_t}
  R(t) = \frac{g^2}{8\pi^2} \int_{x_{\mathrm{min}}(t,E^2)}^{x_{\mathrm{max}}(t,E^2)} dx\, \frac{1}{t} P_{\chi\to\chi} (x,t)\;,
\end{align}
where the Altarelli-Parisi splitting function~\cite{ALTARELLI1977298} with massive partons is given by~\cite{Catani:2002hc}~\footnote{
When we specify the mass origin of the dark sector, we need to take dark fermion helicities into account, leading to additional model-dependent contributions to the splitting kernel (see \cite{Chen:2018uii} for a more detailed discussion).
}
\begin{align}
  P_{\chi\to\chi} (x, t) = \frac{1+x^2}{1-x} - \frac{2(m_\chi^2 + m_{A'}^2)}{t}\;.
  \label{eq:splitting}
\end{align}
The Sudakov factor, which expresses the no-splitting probability between $t_1 < t < t_0$, is obtained as
\begin{align}
  \Delta (t_0, t_1) = \exp \left[ - \int_{t_1}^{t_0} dt\, R(t) \right]\;.
\end{align}

For the analytical treatment of the multi-emission processes, we use the approximation to replace $E$ for each splitting step by the initial parton energy scale $E_0$.  This approximation is valid if each emission is soft enough, which is expected to be true for most cases.
Under this approximation, the probability that $\chi$ with initial energy $E_0$ ends up with only one dark photon in the final state is given by
\begin{align}
  p_1 = \int_{\tmin}^{\tmax} dt\,
  \Delta \left( \tmin, t \right) R(t) \Delta \left( t, \tmax \right)\;.
\end{align}
The above expression can be simplified using
\begin{align}
  N \equiv \int_{\tmin}^{\tmax} dt\, R(t)\;,
\end{align}
as
\begin{align}
  p_1 = N e^{-N}\;.
\end{align}
In general, the probability of obtaining $n$ dark photons in the final state is given by the Poisson distribution
\begin{align}
  p_n = \frac{N^n}{n!} e^{-N}\;,
\end{align}
so that $\sum_n p_n = 1$ as desired.
Note that the expectation value of the number of dark photons is
\begin{align}
  \Braket{n} \equiv \sum_n n p_n = N\;.
\end{align}

\section{Quantum interference effects}
\label{sec:interference}
\setcounter{equation}{0}

Next, let us consider the Lagrangian with $N_f$ flavors of the dark fermions,
\begin{align}
  \mathcal{L}_{\mathrm{dark}} =
  \sum_i \bar{\chi}_i (i\slashed{\partial} - m_{\chi_i}) \chi_i
  + \sum_{i,j} ig_{ij} \bar{\chi}_i \slashed{A}' \chi_j
  - \frac{1}{4} F'_{\mu\nu} F'^{\mu\nu} - \frac{1}{2} m_{A'}^2 A'_\mu A'^\mu\;,
\end{align}
where $i,j=1,\dots N_f$ are the flavor indices.
In this multi-flavor case, the emission process can be affected by the quantum interference between two diagrams that have different flavors of $\chi_i$ propagating as an intermediate state.
Below, we analytically quantify such effects in some specific choice of model parameters.

For simplicity of discussion, we consider the case with $m_{\chi_i} = m_\chi$ and $g_{ij} = g$ for any $i,j$.
In this case, the probability of the splitting process $\chi_i \to \chi_j A'$ is given by $R_{ij}(t) = R(t)$, 
using the functions defined previously in \eqref{R_t}. 
The Sudakov factor for the flavor $i$ is then calculated as
\begin{align}
  \Delta_i (t_0, t_1) = \exp \left[ - \sum_j \int_{t_0}^{t_1} dt\, R_{ij}(t) \right]
  = \Delta (t_0, t_1)^{N_f}\;.
\end{align}

The classical expectation value of the number of emitted dark photons is estimated in the same way as in the previous section using the classical parton shower.
The probability of the $\chi_i \to \chi_j$ process with only one or less dark photon in the final state is given by
\begin{align}
  \hat{p}_0^{(i\to j)} &= e^{-N_f N} \delta_{ij}\;, \label{eq:p0cc} \\
  \hat{p}_1^{(i\to j)} &= \int_{\tmin}^{\tmax} dt\;
  \Delta_i \left( \tmin, t \right) R_{ij}(t) \Delta_j \left( t, \tmax \right) = N e^{-N_f N}\;,
\end{align}
and similarly for multi-dark-photon processes
\begin{align}
  \hat{p}_n^{(i\to j)} = \frac{N_f^{n-1}}{n!} N^n e^{-N_f N}
  ~~~(n\geq 1)\;,
  \label{eq:nemission-classical}
\end{align}
where the factor $N_f^{n-1}$ corresponds to the choice of the flavors of the $n-1$ intermediate fermions.
These probabilities satisfy
\begin{align}
  \sum_{j} \sum_n \hat{p}_n^{(i\to j)} &= 1\;, \\
  \widehat{\Braket{n}} \equiv \sum_j \sum_n n \hat{p}_n^{(i\to j)} &= N_f N\;, \label{classical_expectation_value}
\end{align}
where the second equation shows the expected number of emissions in the inclusive process $\chi_i \to \text{all}$.

However, the above estimate misses quantum interference effects between different flavors of intermediate fermions.
As an example, the missing contributions to the two dark photon emission process are
\begin{equation}
  \sum_{k\neq \ell}~
  \feynmandiagram[horizontal=i to f, inline=(c)] {
    i -- [fermion, edge label=\(\chi_i\)] a
      -- [fermion, edge label=\(\chi_k\)] b
      -- [fermion, edge label=\(\chi_j\)] f,
    a -- [photon] c,
    b -- [photon] d,
  };
  \left[
  \feynmandiagram[horizontal=i to f, inline=(c)] {
    i -- [fermion, edge label=\(\chi_i\)] a
      -- [fermion, edge label=\(\chi_\ell\)] b
      -- [fermion, edge label=\(\chi_j\)] f,
    a -- [photon] c,
    b -- [photon] d,
  };
  \right]^{*}+\text{c.c.}\;,
\end{equation}
in the amplitude squared.
The main reason why the classical parton shower calculation misses these contributions is that it takes the classical probabilistic approach.
The flavor index of the intermediate dark fermion is considered at the cross section level, which corresponds to consider the sum over the flavor index of the amplitude squared instead of the square of the sum of the amplitudes.

\begin{table}[t]
  \renewcommand{\arraystretch}{1.3}
  \begin{tabular}{cc|cccc}
    & & \multicolumn{4}{c}{\# of emissions} \\
    & & 0 & 1 & 2 & 3 \\ \hline
    \multirow{4}{*}{\rotatebox{90}{\# of loops}} & 0 & $1$ & $N_f N$ & $N_f^3 N^2/2$ & $N_f^5 N^3/6$ \\
    & 1 & $-N_f N$ & $-N_f^3 N^2$ & $-N_f^5 N^3/2$ & \dots \\
    & 2 & $N_f^3 N^2/2$ & $N_f^5 N^3/2$ & \dots & \\
    & \vdots & \vdots &&& \\ \hline
    & Total & $\left( N_f - 1 + e^{-N_f^2 N} \right) /N_f$ & \dots & &
  \end{tabular}
  \centering
  \caption{
    Contributions to the cross sections normalized by the fixed-order tree-level cross section without dark photon emission.
    The $k$-loop contributions should be understood as those from the interference between the $k$-loop and the tree diagrams with the same number of emissions.
  }
  \label{tab:quantum-PS}
\end{table}

Fortunately, in our simplified setup, it is straightforward to include the missing pieces due to the universality of dark fermions.
For processes with $n$ emissions, $N_f^{n-1}$ terms in the amplitude squared are included in the classical result \eqref{eq:nemission-classical}, while $N_f^{2(n-1)}-N_f^{n-1}$ are missed, where each term gives exactly the same contribution to the cross section.\footnote{
Here, we consider the cross section of any production process of a dark fermion and several SM particles associated with dark photon emissions.
}
This determines the tree-level contributions to the cross section for an arbitrary number of dark photon emissions as summarized in the first line of Table~\ref{tab:quantum-PS}.
Each element of the table is the contribution to the corresponding cross section normalized by the fixed-order tree-level cross section without dark photon emission.
All the other lines of the table shows contributions of the interference between the $k$-loop and the tree-level diagrams of the $n$-emission process, which are the unique loop-level contributions at the leading logarithmic level (see, e.g., \cite{DOKSHITZER1980269}).
From the unitarity, the order-by-order cancellation of the IR divergence, and thus of the large logarithms, are ensured.
This allows us to fill all the elements of Table~\ref{tab:quantum-PS}.
In the final line, we show the normalized cross sections, or the probability distribution against the number of emissions.
We can obtain the analytic formulas
\begin{align}
  p_0^{(i\to j)} &= \left( \frac{N_f - 1}{N_f} + \frac{1}{N_f} e^{-N_f^2 N} \right) \delta_{ij}\;, \label{eq:p0ij} \\
  p_n^{(i\to j)} &= \frac{1}{N_f^2} \frac{1}{n!} (N_f^2 N)^{n} e^{-N_f^2 N}
  ~~~
  (n\geq 1)\;.
\end{align}
Note that the expression \eqref{eq:p0ij} has a physical interpretation as follows.
Since the eigenvalues of the gauge coupling matrix $g_{ij}$ are $\{ N_f g, 0, 0, \dots \}$, we find a initial state with vanishing gauge eigenvalue with the probability $(N_f-1)/N_f$, which corresponds to the first term.
Similarly, the second term is obtained as a product of the probability $1/N_f$ with which we find a initial state with non-vanishing gauge eigenvalue $N_f g$ and the corresponding non-emission probability $e^{-N_f^2 N}$.
Note also that the inclusive probabilities $p_n$ and $\hat{p}_n$ defined as
\begin{align}
    p_n \equiv \sum_{j} p_n^{(i\to j)}\;,
    \label{eq:inclusive}
\end{align}
and similar for $\hat{p}_n$, are independent of the initial flavor $i$ for any $n\geq 0$ due to the flavor universality.
The correct expectation value of emissions in the inclusive process is calculated as
\begin{align}
  \Braket{n} \equiv \sum_n n p_n = N_f N\;,
\end{align}
which coincides with the classical result \eqref{classical_expectation_value}. 

\begin{figure}[t]
  \centering
  \includegraphics[width=0.48\hsize]{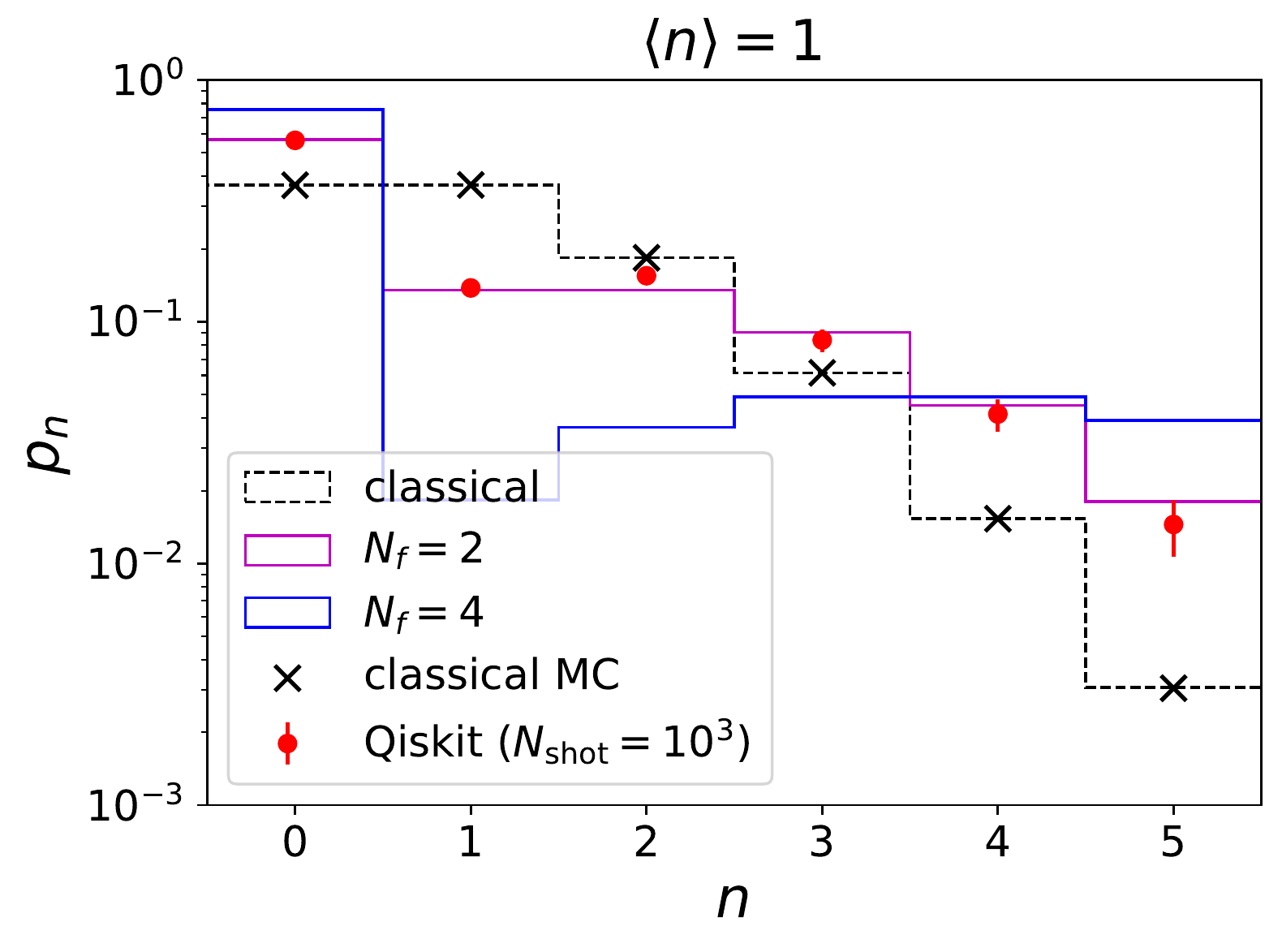}
  \includegraphics[width=0.48\hsize]{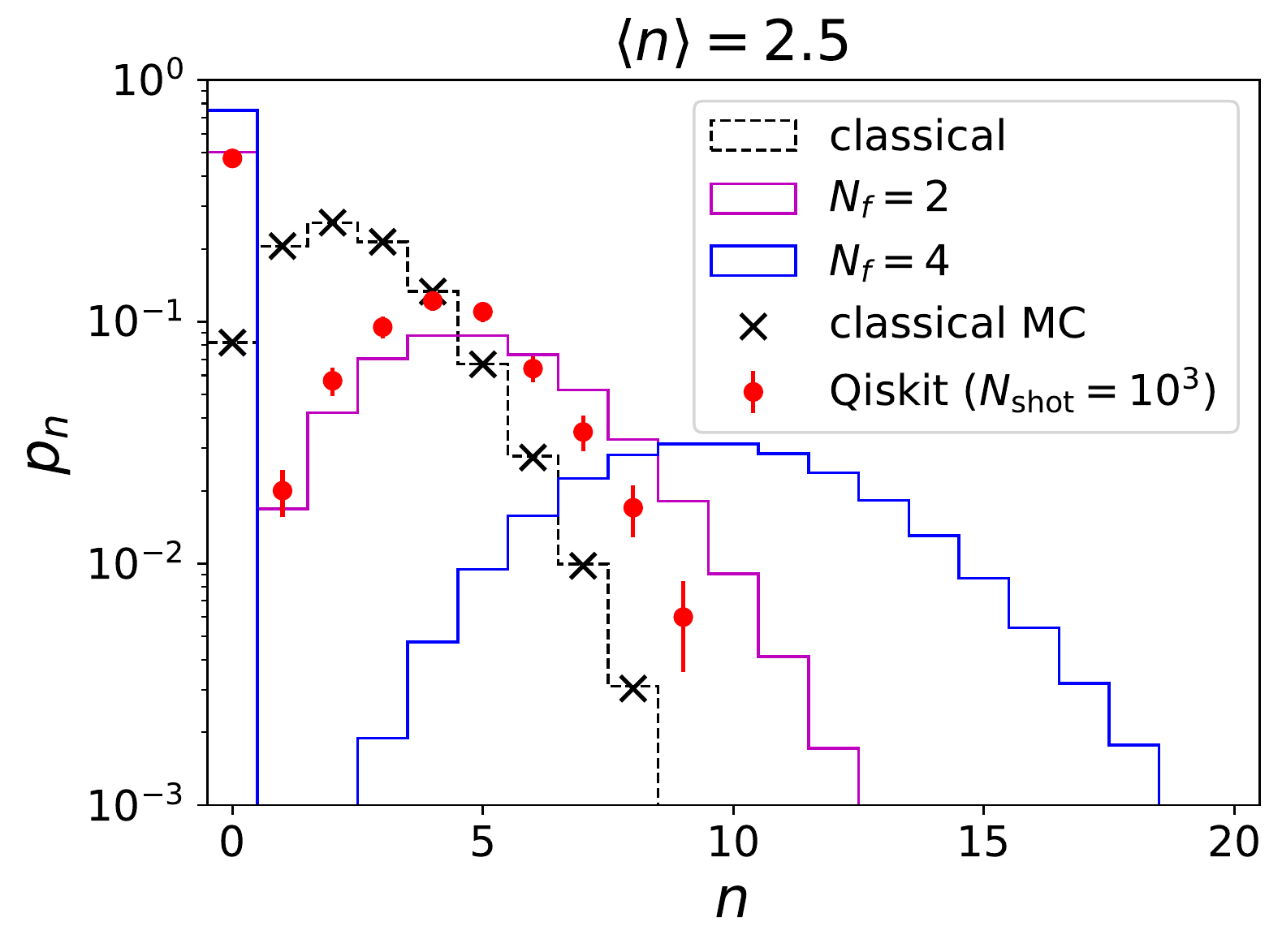}
  \caption{
  Histograms of the probability distributions $\hat{p}_n$ and $p_n$ as functions of $n$ defined in \eqref{eq:inclusive}.
  The left (right) panel shows the case with $\Braket{n} = 1$ ($2.5$).
  Also shown by crosses and circles are the numerical results described in Sec.~\ref{sec:simulation}.
  Note that $\alpha_1' \simeq 0.14$ $(0.35)$ corresponds to $\Braket{n} = 1$ ($2.5$) with the same choice of energy and masses as in Sec.~\ref{sec:simulation}.
  }
  \label{fig:quantum-interference}
\end{figure}

Although $\widehat{\Braket{n}} = \Braket{n}$, the probability distributions as a function of the number of emissions $n$ clearly differ between classical and quantum calculations.
In Fig.~\ref{fig:quantum-interference}, we show $\hat{p}_n$ (black dashed) and $p_n$ (solid) as a function of $n$.
The left (right) panel shows results for $\Braket{n} = N_f N = 1$ (2.5), while different colors discriminate different choices of $N_f = 2$ (red), $3$ (green), and $4$ (blue).
Note that the classical calculation gives the same results shown by the black dashed lines irrespective of the choice of $N_f$.
The figure indicates that the peak position of the probability distribution is proportional to $N_f$ and models with larger $N_f$ tend to have a higher probability with large $n$.
At the same time, however, $p_{n=0}$ increases for larger $N_f$ so that the expectation value $\Braket{n}$ does not change with $N_f$.\footnote{
Note that the significant enhancement of $p_{n=0}$ for a large $N_f$ is an artifact of our special choice of parameters, where $N_f-1$ out of $N_f$ flavors are chargeless when the gauge interaction is diagonalized.
For a general choice of parameters, we can see from the numerical calculation described below that the modification is more moderate, though the qualitative behavior is the same.
}

In the figure, we also show the numerical results of the classical parton shower calculation with black crosses.
We implement the flavor index of dark fermions in the classical parton shower and simulate the splitting according to the probability distribution $R_{ij}(t) = R(t)$.
The black crosses show results for $N_f = 2$ with $N_f N = 1$ $(2.5)$ in the left (right) panel, which agree with the theoretical expectation shown by the black dashed line.
We numerically checked that the results does not depend on the choice of $N_f$ within the statistical errors when $N_f N$ is fixed.

\section{Quantum simulations}
\label{sec:simulation}
\setcounter{equation}{0}

Next, we perform quantum simulations of the parton shower to quantify the quantum interference effects for different choices of model parameters.
Since our simulation treats a superposition of states with and without radiations at each step, the kinematic variables such as the energy of the initial parton at each step are intractable.
Thus, we resort to the approximation $E\simeq E_0$ discussed above to perform a meaningful calculation.
The effect of this approximation on the expectation value of emissions is typically of $O(1)\%$ depending on the model parameters~\cite{Buschmann:2015awa}.

\begin{figure}[t]
  \centering
  \includegraphics[width=0.9\hsize]{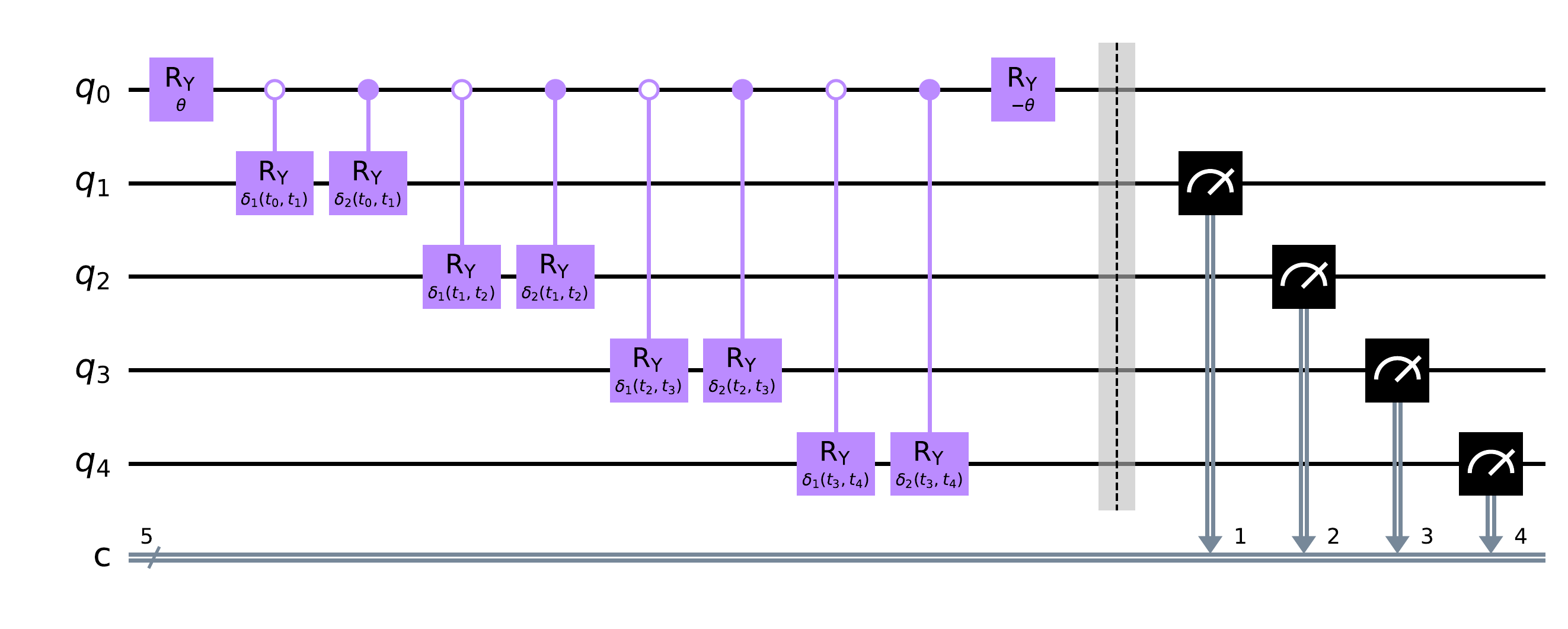}
  \caption{Quantum circuit for parton shower simulation for $N_f=2$ and $\nstep=4$.}
  \label{fig:circuit}
\end{figure}

The configuration of our quantum circuit is the same as the one presented in \cite{Bauer:2019qxa}.
We discretize the virtuality $t$ into $\nstep$ steps with threshold values $t_{0} \equiv t_{\text{max}} > t_1 > \dots > t_{\nstep} \equiv t_{\text{min}}$.
Also, we focus on the case with $N_f=2$, which is the easiest to tackle with a limited number of qubits and gates.
Then, we need $\nstep+1$ qubits; the first qubit $\ket{q_0}$ represents the state vector of the dark fermion, while the other $\nstep$ qubits $\ket{q_i}$ ($i=1,\dots,\nstep$) preserve whether the dark photon emission occurs during $t_{i} < t < t_{i-1}$.
The relevant circuit for $\nstep=4$ is shown in Fig.~\ref{fig:circuit} as an example.

The first $R_Y$ gate works as a rotation by an angle $\theta$, and this changes the basis of fermions into the one in which the gauge interaction is diagonal.
The rotation angle $\theta$ is defined so that $U_\theta G U_\theta^\dagger = \text{diag} (g_1', g_2')$ with
\begin{align}
  U_\theta \equiv \begin{pmatrix}
    \cos \frac{\theta}{2} & -\sin \frac{\theta}{2} \\
    \sin \frac{\theta}{2} & \cos \frac{\theta}{2}
  \end{pmatrix}
  ~~;~~
  G \equiv \begin{pmatrix}
    g_{11} & g_{12} \\
    g_{21} & g_{22}
  \end{pmatrix}\;.
\end{align}
The controlled $R_Y$ gates rotate qubits $\ket{q_i}$ ($i=1,\dots,\nstep$) so that the probability of finding the state $\ket{q_i}=\ket{1}$ at the measurement equals to the corresponding dark photon emission probability.
Then, the rotation angle $\delta_f (t_{i-1}, t_i)$ ($f=1,2$) are determined through the condition
\begin{align}
  \tan \left( \frac{\delta_f (t_{i-1}, t_i)}{2} \right) =
  \sqrt{\frac{1 - \Delta_f' (t_{i-1}, t_i)}{\Delta_f' (t_{i-1}, t_i)}}\;,
\end{align}
where $\Delta_f'$ are Sudakov factors calculated with the eigenvalues of the gauge coupling $g_f'$.

There are two reasons why we can treat each emission independently for a general choice of the gauge coupling matrix $G$ as shown in Fig.~\ref{fig:circuit}.
Firstly, the approximation $E\simeq E_0$ ensures that the emission probability can be calculated irrespective of whether the emission has occurred in the previous steps or not.
Secondly, because we work in the eigenbasis of the gauge couping matrix during the emission steps, each emission does not change the fermion state stored in the qubit $q_0$.

\begin{figure}[t]
  \centering
  \includegraphics[width=0.8\hsize]{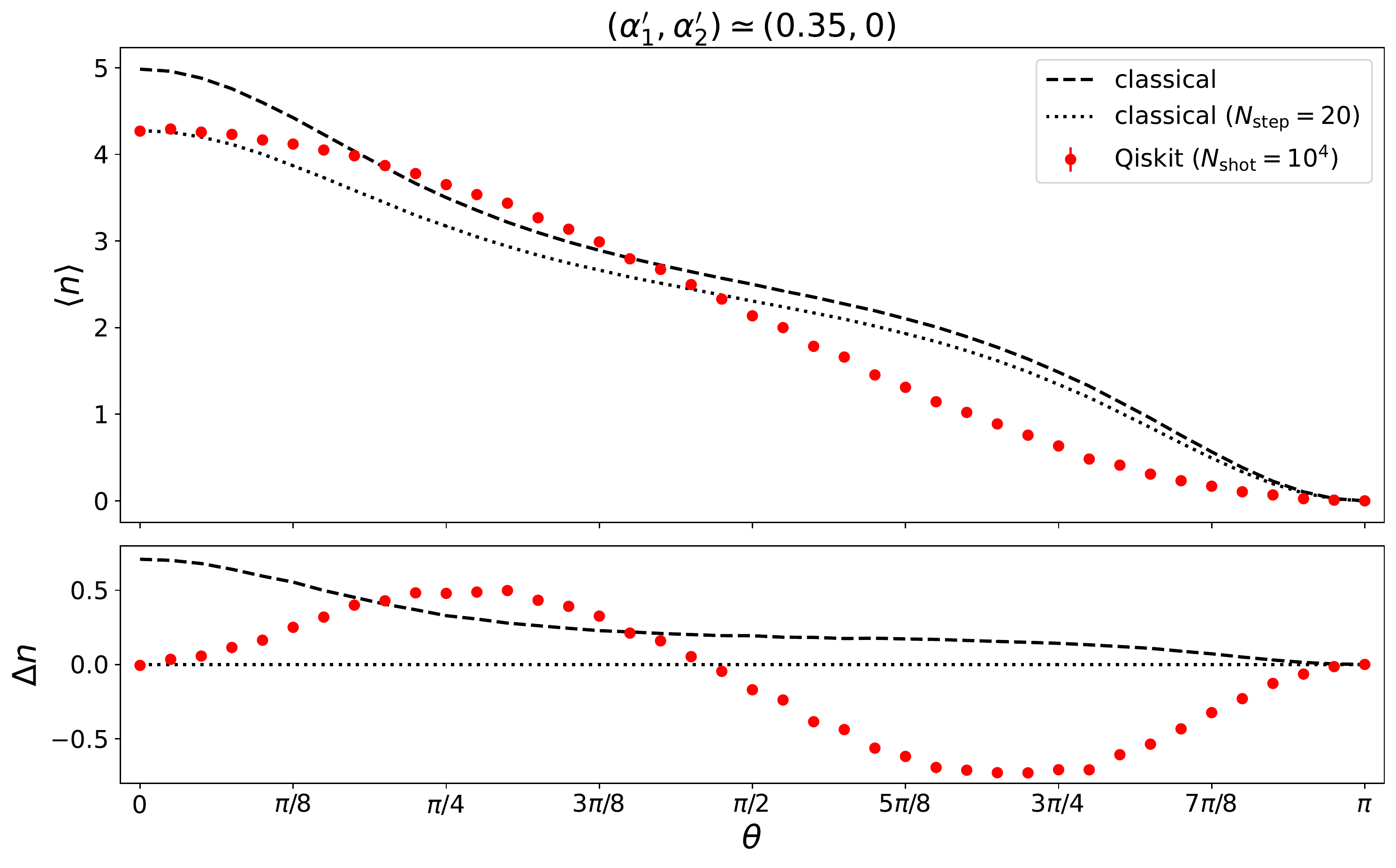}
  \caption{
  Simulation results for $(\alpha_1', \alpha_2') \simeq (0.35, 0)$ as a function of $\theta$, which parametrizes the initial state.
  Average numbers of emissions in the Qiskit (classical) calculation are shown by red circles (black lines).
  $\nshot=10^3$ and $\nstep=20$ are used for the Qiskit calculation, while $\nshot=10^6$ and $\nstep$ is sufficiently large ($\nstep=20$) for the classical calculation shown by the black dashed (dotted) line.
  \textit{Top:} Expectation number of emissions $\Braket{n}$.
  \textit{Bottom:} Change in the expected number $\Delta n$ compared with the black dotted line.
  }
  \label{fig:nexp-vs-theta}
\end{figure}

In Fig.~\ref{fig:nexp-vs-theta}, we compare the results of the quantum and classical parton shower simulations.
For the quantum results, we use a simulator of the quantum computer through the IBM Qiskit \cite{Qiskit}.
As for model parameters, we use fixed values of $(\alpha_1', \alpha_2') \simeq (0.35, 0)$ with $\alpha_f' \equiv g_f'^2 / 4\pi$ ($f=1,2$), $E_0 = 500\,\mathrm{GeV}$, $m_\chi=0.4\,\mathrm{GeV}$, and $m_{A'} = 0.4\,\mathrm{GeV}$, which gives $N_f N=2.5$ in the flavor universal basis.
The corresponding gauge coupling matrix $G$ can be calculated as a function of $\theta$.
We fix the initial state to be $\ket{q_0}=\ket{0}$ and see how the emission process depends on $\theta$.
The classical calculations are shown by the black dashed lines, for which 
we use $\nshot=10^6$ and a sufficiently large number of steps $\nstep\sim 10^3$.
The Qiskit calculations are shown by the red circles, with error bars from statistical uncertainties.
Here we used $\nshot=10^3$ and $\nstep=20$, which are within the reach of the IBM Quantum Experience \cite{IBMQ} today.
Due to the limitation in the number of qubits, the discretization of the integration variable is much more coarse 
in the quantum calculations than in the classical calculations,
and this increases numerical discretization errors for the former.
For comparison, we also show the classical results with $\nstep=20$ by the black dotted lines.
In the top panel, we plot the expectation values of the number of emissions $\Braket{n}$, while in the bottom panel we plot the difference in expectation values $\Delta n$ compared to the results shown by the black dotted line.

The $\sim 10\,\%$ gap between the black dashed and dotted lines represent the numerical errors sourced from the use of small step number $\nstep=20$.
These errors are expected in the Qiskit results as well as those shown by the black dotted lines.
To reduce these errors for the quantum simulation, it is needed to use more qubits to increase $\nstep$.
By comparing the Qiskit results with the black dotted lines, it can be seen that the quantum and classical calculation results agree with each other when $\theta\simeq 0$, $\pi/2$, and $\pi$.\footnote{The small deviation between the quantum and classical calculation results at $\theta = \pi/2$ is due to the difference in the probability distribution shown in Fig.~\ref{fig:quantum-interference}.
The numerical discretization errors cause the peak shift of the probability distribution, whose effect on the expected number of emissions weakly depends on the original distribution.}
The first and the last choices correspond to the models without flavor mixing, while the second choice corresponds to the flavor universal case discussed in the previous section.
On the other hand, we can see that the expectation value $\Braket{n}$ changes due to the quantum interference effect for any $\theta$ other than these specific values.
The sign of the quantum interference contribution depends on the choice of $\theta$.

\begin{figure}
    \centering
    \includegraphics[width=0.48\hsize]{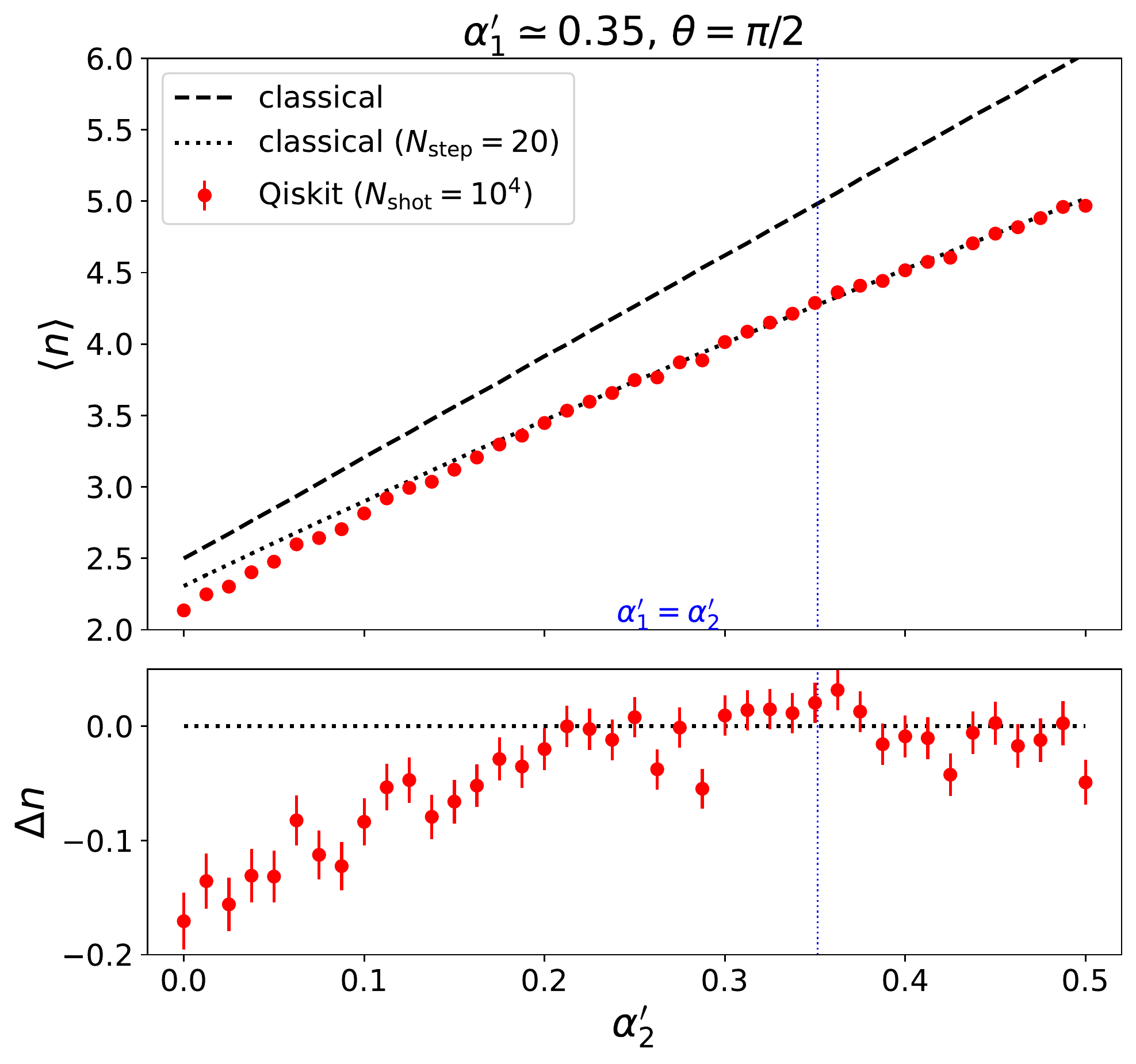}
    \includegraphics[width=0.48\hsize]{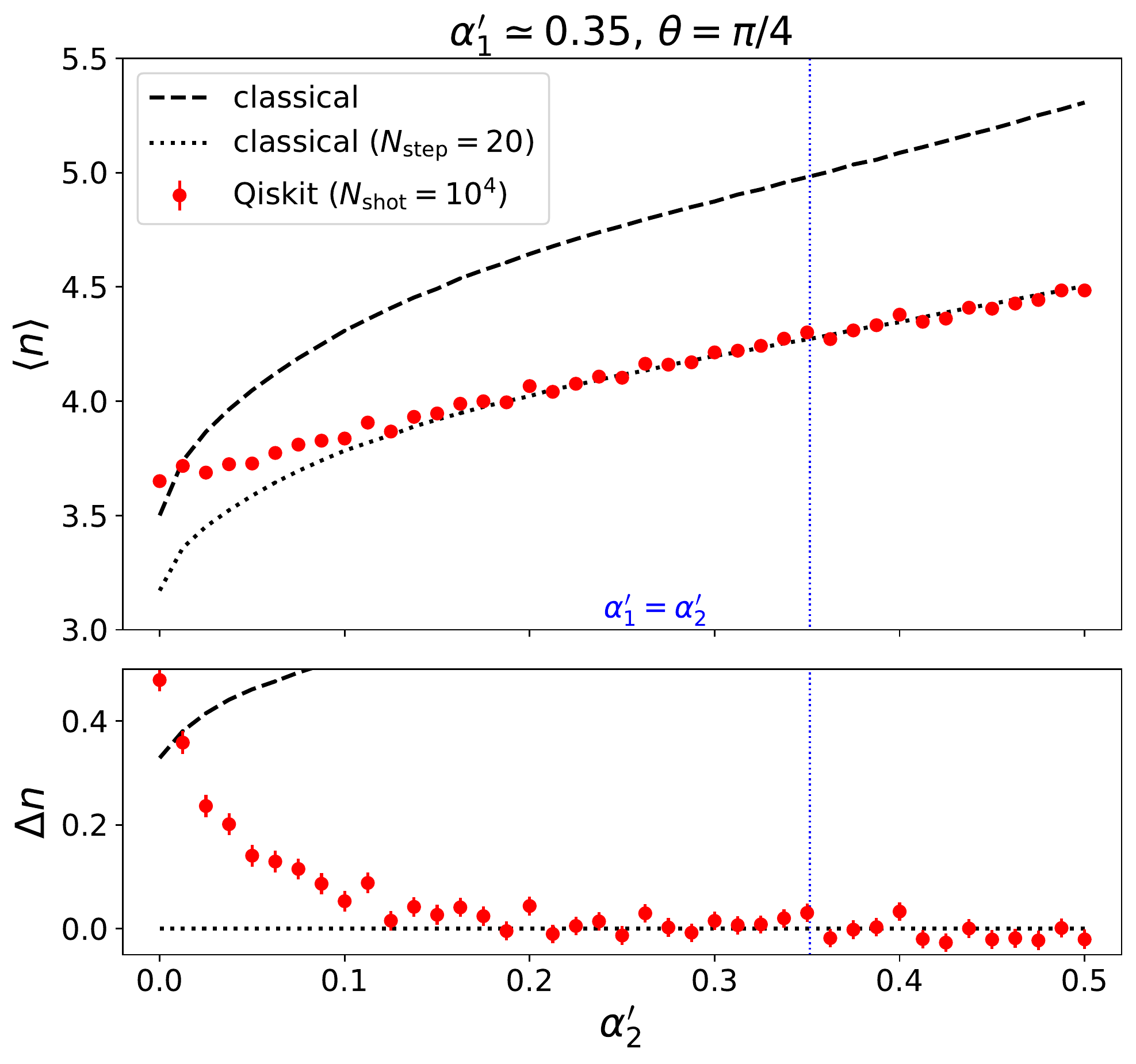}
    \caption{
    Simulation results for $\alpha_1' \simeq 0.35$ as a function of $\alpha_2'$.
    $\theta=\pi/2$ ($\pi/4$) is used for the left (right) panel.
    The simulation setup and the plot convention is the same as Fig.~\ref{fig:nexp-vs-theta}.
    }
    \label{fig:nexp-vs-alpha}
\end{figure}

In Fig.~\ref{fig:nexp-vs-alpha}, we show similar plots with varying $\alpha_2'$.
The other gauge coupling $\alpha_1' \simeq 0.35$ is fixed and $\theta = \pi/2$ and $\pi/4$ are used in the left and right panels, respectively.
Note that the vertical blue dotted lines show the parameter choice $\alpha_1' = \alpha_2'$, where the flavor mixing is absent irrespective of the choice of $\theta$ and thus $\Delta n = 0$.
As a general rule, the quantum interference effect tends to be more important when displaced from blue lines, i.e., when the ratio $\max (\alpha_1', \alpha_2') / \min (\alpha_1', \alpha_2')$ is large.
The unique exception is the flavor universal case discussed above.

\begin{figure}
    \centering
    \includegraphics[width=0.6\hsize]{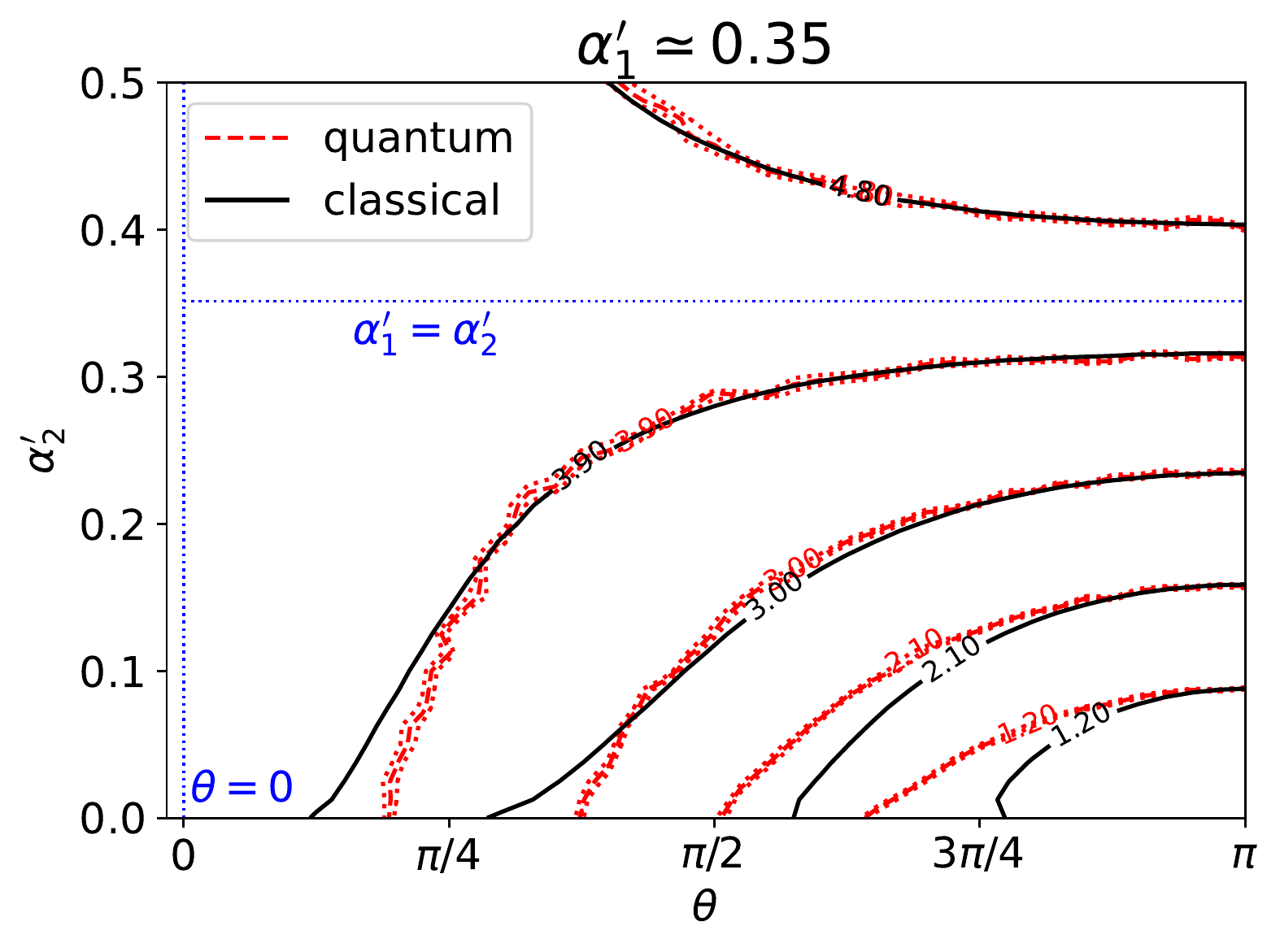}
    \caption{
    Contour plot of $\Braket{n}$ in the $\theta$ vs. $\alpha_2'$ plane.
    Results of the classical (quantum) parton shower are shown by the black solid (red dashed) line.
    The quantum results are accompanied by the statistical errors shown by red dotted lines.
    }
    \label{fig:nexp-2D}
\end{figure}

We also show the contour plot of $\Braket{n}$ in the $(\theta, \alpha_2')$ plane in Fig.~\ref{fig:nexp-2D}.
The black solid and the red dashed lines correspond to the results of the classical and quantum simulations, respectively.
The red dotted lines show the statistical errors of the quantum calculations.
On the blue dotted lines that represent $\theta=0$ and $\alpha_1' = \alpha_2'$, there is no flavor mixing and the expectation value is $\Braket{n}=5$, though the discretization error degrades the value to $\Braket{n}\simeq 4.3$ in our calculation.
We can see that, in a large part of the parameter space, the effect of the quantum interference is small and comparable to the statistical errors associated with the current setup.

To this point, we have discussed the effects of the quantum interference on the expectation value $\Braket{n}$.
Let us next discuss probability distributions of the number of dark photon emissions. This contains more information than the mere expectation value---in the previous section, we have already seen for the flavor universal case
that the probability distribution is affected by quantum effects while $\Braket{n}$ remains unaffected.
We plot the simulated probability distribution in Fig.~\ref{fig:quantum-interference} for the Qiskit (red circle) and the classical (black cross) simulations.
From the left panel, we find that both results agree with the analytic results obtained in Sec.~\ref{sec:interference} within the statistical errors.
Note that the peak position of the Qiskit result in the right panel is slightly shifted to the lower $n$ direction compared with the analytical result.
This is a result of the discretization error with $\nstep=20$, which is not enough to simulate the events with $n\sim O(5)$.
In this case, there should be a non-negligible probability that more than one emissions occur within a step $t_q < t < t_{q-1}$, which is not taken care of by the current setup.

\begin{figure}
    \centering
    \includegraphics[width=0.48\hsize]{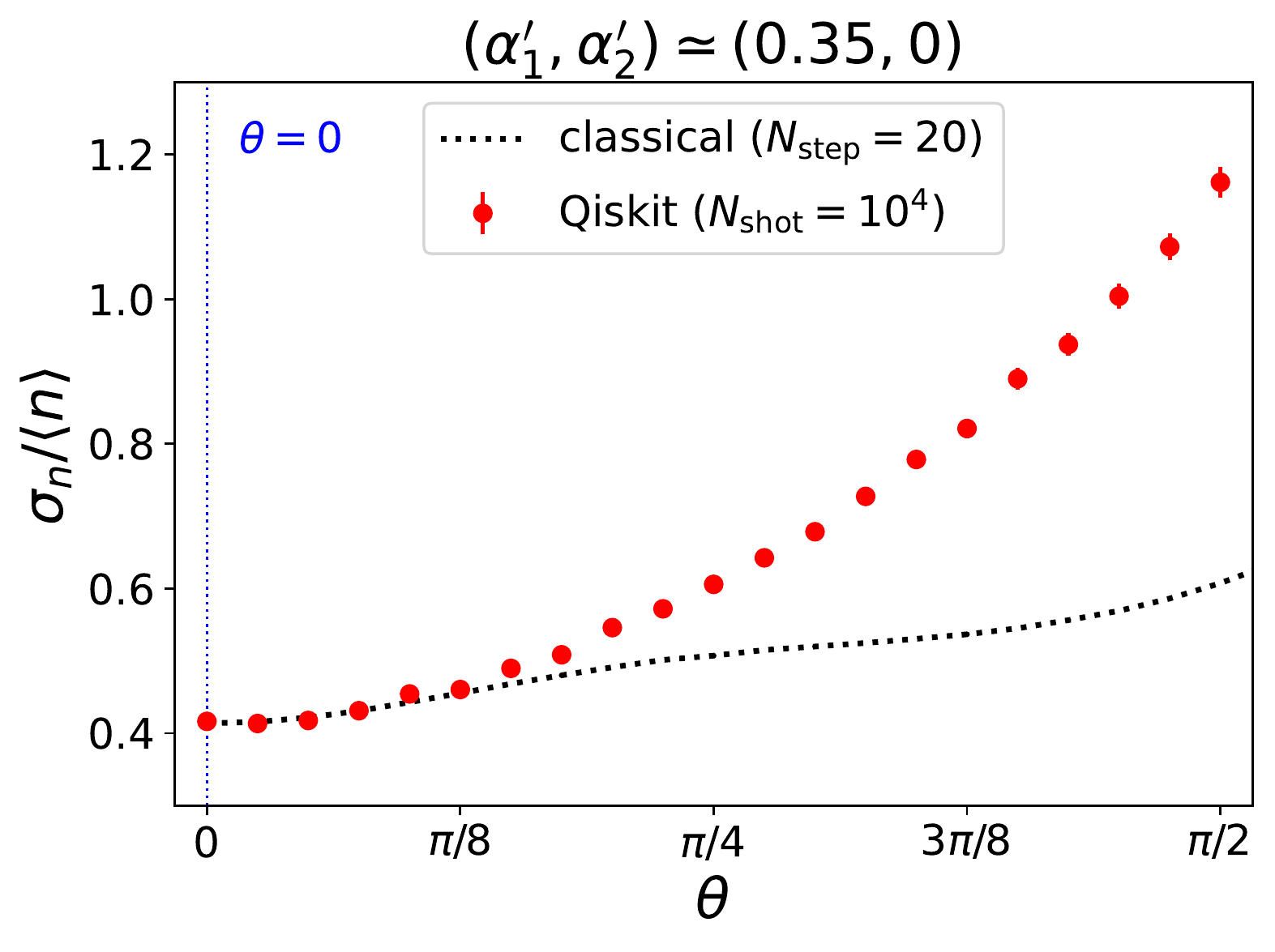}
    \includegraphics[width=0.48\hsize]{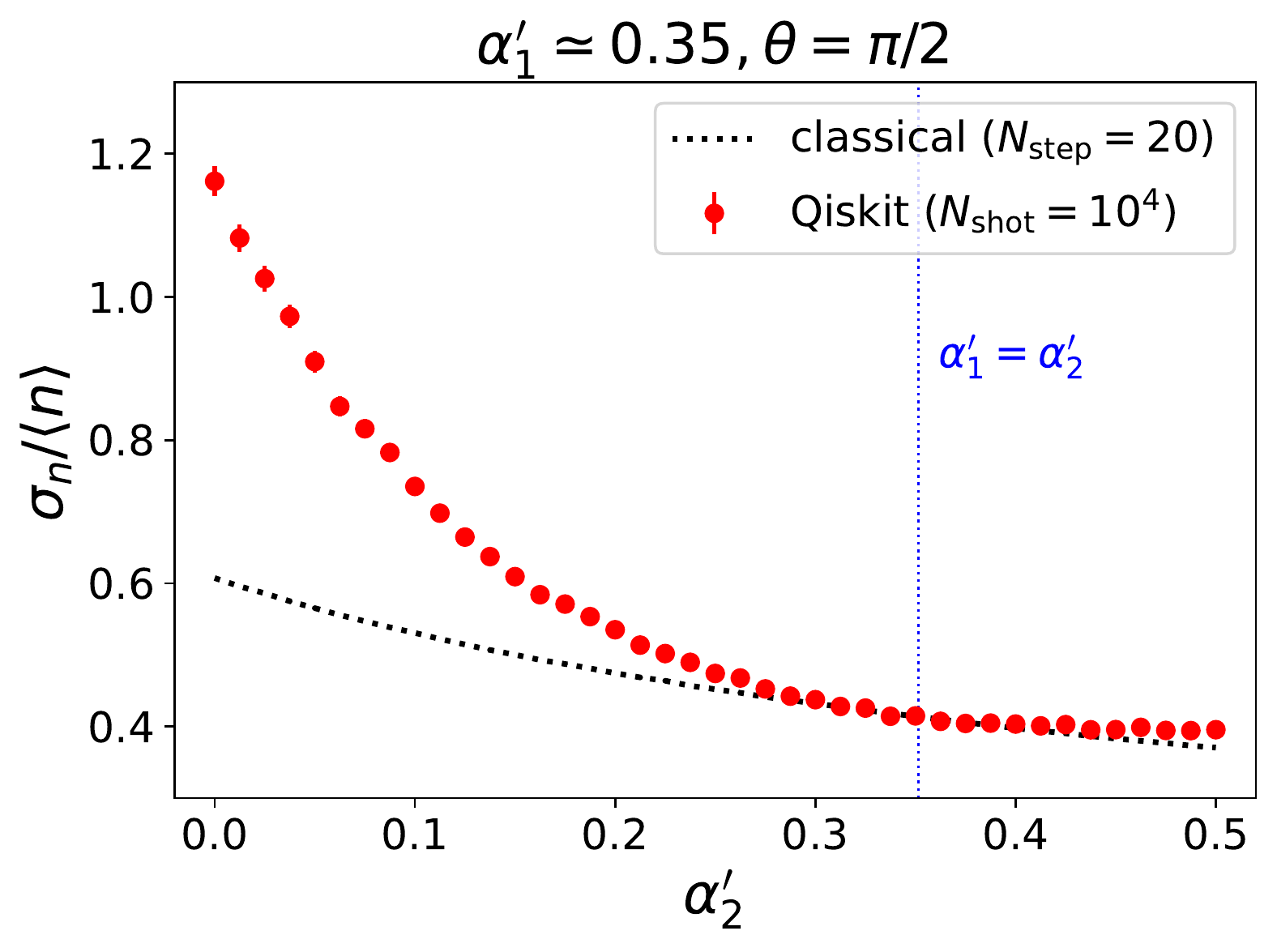}
    \caption{
    \textit{Left:} Plot of $\sigma_n / \langle n \rangle$ as a function of $\theta$ with fixed $(\alpha_1', \alpha_2') \simeq (0.35, 0)$.
    \textit{Right:} Plot of $\sigma_n / \langle n \rangle$ as a function of $\alpha_2'$ with fixed $\alpha_1' \simeq 0.35$ and $\theta=\pi/2$.
    In both panels, the plot convention is the same as in Fig.~\ref{fig:nexp-vs-theta}.
    }
    \label{fig:sigman-1D}
\end{figure}

In order to quantify the size of the quantum interference effect, we use the variance $\sigma_n^2\equiv \langle n^2 \rangle - \langle n \rangle^2$ of the number of emissions.
In Fig.~\ref{fig:sigman-1D}, we show a plot of $\sigma_n / \langle n \rangle$ for various choices of the parameters.
In the left panel, we show $\sigma_n / \langle n \rangle$ as a function of $\theta$ with fixing $(\alpha_1', \alpha_2') \simeq (0.35, 0)$.
The quantum and classical simulations agree with each other for $\theta=0$, where the flavor mixing is absent, while the quantum calculation gives a larger normalized variance for $\theta>0$.
This result is consistent with the observation so far; the quantum interference effect is larger for a process with more emissions, $p_{n+1}/p_n > \hat{p}_{n+1}/\hat{p}_n$, which enhances $\langle n^2 \rangle$ more efficiently than $\langle n \rangle^2$.
Note that the quantum interference effect on $\sigma_n$ is important also for $\theta\simeq \pi/2$, where $\Braket{n}$ is unaffected.
In the right panel, we show a plot as a function of $\alpha_2'$ with fixed $\alpha_1' \simeq 0.35$ and $\theta = \pi/2$.
Similarly to the left panel, we can see that the normalized variance takes a larger value in the quantum calculation for most values of $\alpha_2'$.
The unique exception is the case $\alpha_1' = \alpha_2'$, where the flavor mixing is absent.

\begin{figure}
    \centering
    \includegraphics[width=0.6\hsize]{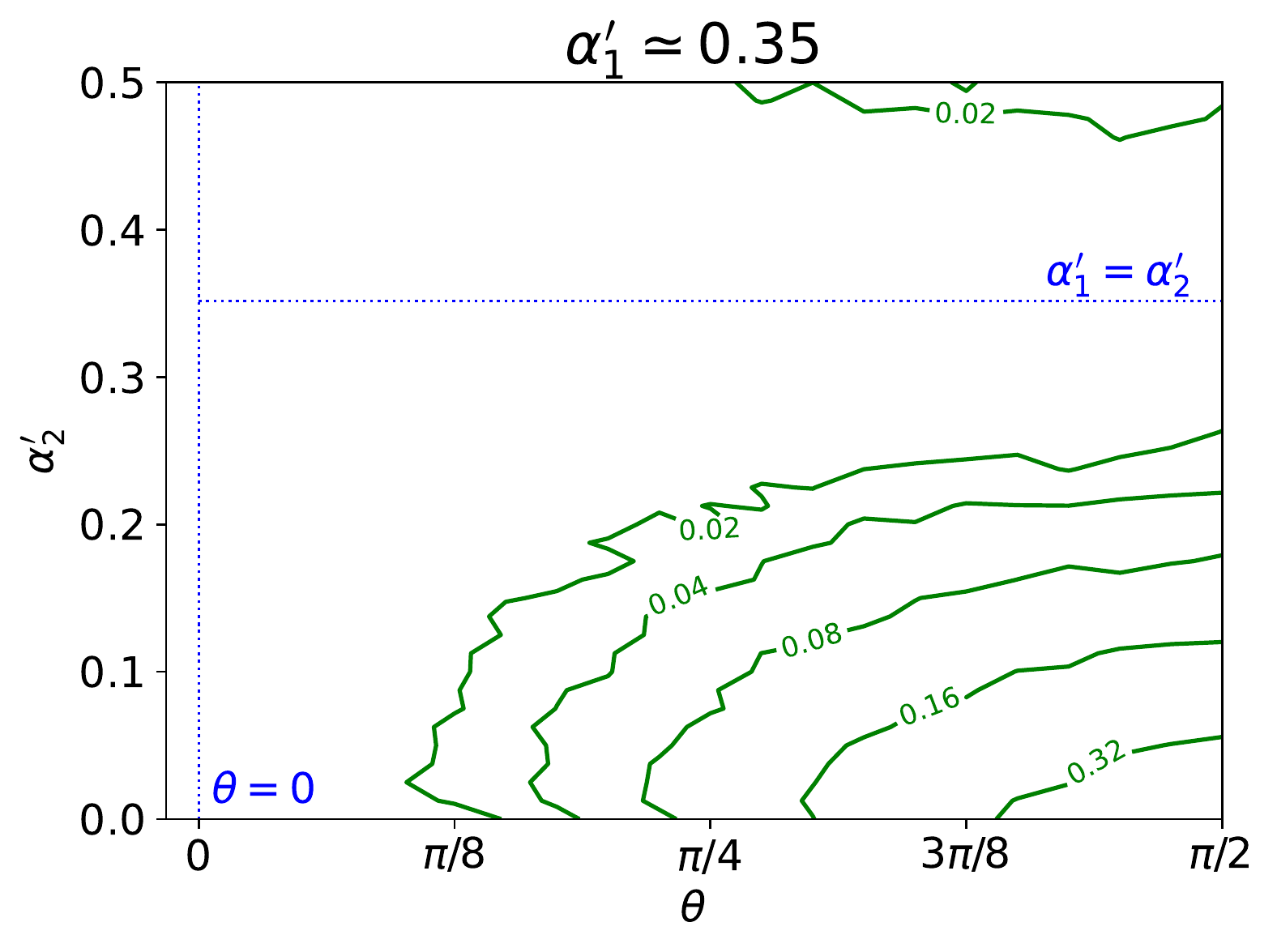}
    \caption{
    Contour plot of $\Delta (\sigma_n / \langle n \rangle)$ in the $\theta$ vs. $\alpha_2'$ plane.
    }
    \label{fig:sigman-2D}
\end{figure}

Finally, we show the difference in the normalized variance between the quantum and classical simulations, $\Delta (\sigma_n / \langle n \rangle)$, as a benchmark of the quantum interference effects.
We show the contour plot of $\Delta (\sigma_n / \langle n \rangle)$ in the $\theta$ vs.\ $\alpha_2'$ plane in Fig.~\ref{fig:sigman-2D}.
Contours are jagged simply because of the statistical errors of the Qiskit calculation.
Note that the blue dotted lines give the minimum of $\Delta (\sigma_n / \langle n \rangle) = 0$ due to the absence of the flavor mixings. It is easy to see that the interference effect is more important for parameter choices displaced from these lines, i.e., when $\theta \simeq \pi/2$ and the ratio $\max (\alpha_1', \alpha_2') / \min (\alpha_1', \alpha_2')$ are large.

\section{Discussion}
\label{sec:discussion}
\setcounter{equation}{0}

If we assume that the SM particles are not charged under the dark $U(1)_D$ symmetry, they could interact with the dark sector particles only through the kinetic mixing between $U(1)_D$ and the SM hypercharge $U(1)_Y$.
This interaction makes it possible to produce dark sector particles at the collider experiments and observe the signal originating from the dark sector shower.
Experimentally, dark photons generated by the dark sector shower are detected through their decay into SM particles.
In this case, the branching ratio of the dark photon is uniquely determined as a function of the mass $m_{A'}$~\cite{Falkowski:2010cm}.
In particular, the leptonic decay $A'\to e^{+}e^{-}$ and $\mu^{+}\mu^{-}$, if allowed, is dominant for $m_{A'}\lesssim 0.4\,\mathrm{GeV}$, while the hadronic decay such as $A'\to \pi^{+}\pi^{-}$ becomes more important for $m_{A'}\gtrsim 0.4\,\mathrm{GeV}$.
The leptonic decay of the dark photon leads to the high-multiplicity clusters of boosted and collimated leptons, which are often called the lepton jets~\cite{Cheung:2009su, Meade:2009rb, Falkowski:2010cm, Falkowski:2010gv}.
This characteristic signal is searched for by the ATLAS collaboration with focusing on the prompt lepton jets~\cite{ATLAS:2012wib} and the displaced ones~\cite{ATLAS:2014fzk}.
In both cases, it is important to have a large number of leptons inside a jet cone to reduce the background events from multi-jet events, the cosmic-ray muons, and so on.

It is thus beneficial to take account of the quantum interference effect, which can enhance the high-multiplicity tail of the number of dark photon emissions $n$ as is shown in Fig.~\ref{fig:quantum-interference}.
To see how the sensitivity of the existing analysis can be improved, we give a rough estimate of the signal acceptance rate.
For example, identification of the prompt muon-jets requires at least four muons inside a jet cone~\cite{ATLAS:2012wib}.
We assume the branching ratios $\Gamma(A'\to e^{+}e^{-})=\Gamma(A'\to \mu^{+}\mu^{-})\simeq 0.5$, which corresponds to $m_{A'}\simeq 0.4\,\mathrm{GeV}$, for simplicity.
If $n=2$ dark photons are emitted and they are well collimated with each other such that all the decay products fall into a single jet cone, the probability of having four muons in the final state is $0.25$.
On the other hand, the corresponding probability is $0.5$ if we have $n=3$ dark photons, and even larger for $n>3$.
As a result, the upper limit on the signal acceptance of a dark sector shower as a muon-jet is $9\%$ ($13\%$) for the setup of $N_f=1$ ($N_f=2$) in the left panel of Fig.~\ref{fig:quantum-interference}.
It is also possible that higher probabilities of $n\geq 3$ lead to a sizable number of events with six muons or more, which opens up a possibility to further reduce the background events and improve the sensitivity.
Finally, observation of lepton jets with various $n$ may provide some information on the number of dark flavors $N_f$ through the distribution of $n$ shown in Fig.~\ref{fig:quantum-interference}.

Several comments on possible future directions are in order.
Throughout our analysis, we have fixed the initial flavor of the dark fermion to be $\ket{q_0}=\ket{0}$ and input the rotation angle $\theta$ by hand.
However, $\theta$ should be uniquely determined if we consider some specific production mechanism of the dark fermions.
In this case, we have to perform a more involved simulation including the production processes.
Second, the dark sector shower becomes much more involved when the decay of the dark photon into dark fermions is kinematically allowed: $m_{A'} > 2m_\chi$.
The quantum simulation of such models, which could be of phenomenological importance, is possible with the quantum circuits introduced in \cite{Bauer:2019qxa}.
Third, it is also important to study the kinematics of the emitted dark photons for phenomenological applications.
As mentioned above, the kinematics is intractable in our current treatment because we consider the superposition of states with and without emission at each step. To further discuss phenomenological implications, it is necessary to improve the treatment of the kinematics in the quantum simulation of the parton showers.
Finally, to further increase $\nstep$ and reduce the sizable discretization errors,
it will become necessary to use real-device quantum computers instead of IBM Qiskit simulators.
It will be interesting to quantitatively estimate the capabilities of current and future quantum computers
for our quantum simulations.

\section*{Acknowledgements}

This work grew out of a quantum-simulation study group 
(organized by Y.~Hidaka, T.~Okuda and MY in 2021), and we would like to thank the participants there for stimulating conversations. 
MY would like to thank BCTP/UC Berkeley for their hospitality during the final stages of this project.
The work of SC was supported by the JSPS KAKENHI Grant (20J00046).
The work of SC was supported by the Director, Office of Science,
Office of High Energy Physics of the U.S. Department of Energy under the Contract No.~DE-AC02-05CH1123.
The work of MY was supported in part by WPI Research Center Initiative, MEXT, Japan, and
by the JSPS Grant-in-Aid for Scientific Research (17KK0087, 19K03820, 19H00689, 20H05850, 20H05860).
The Feynman diagrams in this paper are generated by TikZ-Feynman~\cite{Ellis:2016jkw}.

\bibliographystyle{JHEP}
\bibliography{bib}

\end{document}